\documentclass[12pt]{article}

\begin{document}

\tolerance=5000

\def\be{\begin{equation}}
\def\ee{\end{equation}}
\def\bea{\begin{eqnarray}}
\def\eea{\end{eqnarray}}
\def\nn{\nonumber \\}
\def\e{{\rm e}}

\  \hfill
\begin{minipage}{3.5cm}
YITP-02-69 \\
December 2002 \\
\end{minipage}

\vfill

\begin{center}
{\large\bf Logarithmic corrections to the FRW brane cosmology from
5d Schwarzschild-deSitter black hole }

\vfill

{\sc Shin'ichi NOJIRI}\footnote{nojiri@cc.nda.ac.jp},
{\sc Sergei D. ODINTSOV}$^{\spadesuit}
$\footnote{
odintsov@mail.tomsknet.ru}\\
and {\sc Sachiko OGUSHI}$^{\diamondsuit}
$\footnote{JSPS fellow, ogushi@yukawa.kyoto-u.ac.jp}

\vfill

{\sl Department of Applied Physics \\
National Defence Academy,
Hashirimizu Yokosuka 239-8686, JAPAN}

\vfill

{\sl $\spadesuit$
Lab. for Fundamental Studies,
Tomsk State Pedagogical University,
634041 Tomsk, RUSSIA}

\vfill

{\sl $\diamondsuit$
Yukawa Institute for Theoretical Physics,
Kyoto University, Kyoto 606-8502, JAPAN}

\vfill

{\bf ABSTRACT}

\end{center}

Thermodynamics of 5d SdS black hole is considered.
Thermal fluctuations define the (sub-dominant) logarithmic corrections to
black hole entropy and then to Cardy-Verlinde formula 
and to FRW brane cosmology. We demonstrate that logarithmic terms
(which play the role of effective cosmological constant) 
change the behavior of 4d spherical brane in dS, SdS or Nariai bulk.
In particularly, bounce Universe  occurs or 4d dS brane expands 
to its maximum and then shrinks. The entropy bounds are also modified 
by next-to-leading terms. Out of braneworld context the logarithmic terms 
may suggest slight modification of standard FRW cosmology. 

\vfill


\newpage

\section{Introduction}

The deSitter space always attracts much attention in cosmology and 
gravity. This is caused by several reasons. First of all, according to
the theory of inflationary Universe the very early Universe eventually has
passed deSitter (dS) phase. Second, recent astrophysical data indicate 
that modern Universe is (or will be in future) also in deSitter phase.
Third, dS is very attractive from the theoretical point of view due to 
its highly symmetric nature (like flat space). This is also the reason 
why dS space was frequently considered as candidate for ground 
state in quantum gravity.

According to recent studies the dS quantum gravity should be quite unusual 
theory in many respects \cite{witten}.  In connection with braneworld 
scenario \cite{RS} it is expected that there occurs dS/CFT correspondence
\cite{hull, strominger}. In one of its versions,  dS/CFT correspondence
indicates that properties of 5d classical dS space are related with those
of dual 
CFT living on the four-dimensional boundary (which may be also dS).
Despite the fact that explicit examples of such consistent dual CFTs 
are not constructed yet, one can still get a lot of information from dS/CFT 
correspondence. In particular, starting from five-dimensional 
Schwarzschild-deSitter (SdS) black hole (which should be relevant to the
description of 4d dual CFTs at non-zero temperature) one can easily 
get the Friedmann-Robertson-Walker (FRW) brane cosmology.
The corresponding FRW brane equation may be often written in
so-called Cardy-Verlinde (CV) form \cite{EV}. There was much activity 
recently (see \cite{CVdSCFT, SO}) in the study of FRW brane cosmology in CV
form when bulk space is dS or SdS space and in the corresponding
investigations of
thermodynamical properties of dS black holes.

In the present paper we consider 5d SdS black hole and calculate 
the corresponding thermodynamical quantities. Taking into account 
thermal fluctuations defines the logarithmic corrections to both 
cosmological and black hole entropies. As a result the CV formula 
and FRW brane cosmology receive the (sub-dominant) logarithmic corrections,
in the way similar to FRW brane cosmology in AdS black hole bulk
\cite{lidsey}. It is interesting that such sub-dominant terms slightly
change the entropy bounds appearing in CV formulation. 

The paper is organized as follows. In the next section we find  entropy,
free energy and thermodynamical energy for SdS black hole 
which is the space with two horizons and for Nariai black hole
(when both horizons coincide). Using the logarithmic corrections to
the entropy the corrected CV formula is established. The corresponding 
FRW brane equation with logarithmic terms is found.
Section three is devoted to the qualitative study of FRW brane cosmology
where next-to-leading (logarithmic) terms play the role of small effective
cosmological constant. It is explicitly demonstrated that 4d spherical brane 
behaves in a different way when log-terms are present. In particular, 
bounce Universe may occur or dS brane reaches its maximum and then shrinks.
Its behavior depends also from the choice of bulk: dS, SdS 
or Nariai space.
In section four we consider standard 4d FRW cosmology and 
show that even in this case, due to log-corrections to four-dimensional 
cosmological entropy the FRW equation and CV formula may get modified.
The correction terms may be interpreted as dust. Some summary and outlook 
are given in the last section. General derivation of logarithmic corrections 
to entropy (due to thermal fluctuations) is presented in  the Appendix A.
Penrose diagram of SdS space is drawn in Appendix B.

\section{Logarithmic Corrections to Cardy-Verlinde formula and FRW brane
cosmology
in SdS bulk}

Let us consider the thermodynamics of 
SdS black hole in five dimensions. The SdS black hole is a
constant curvature solution of the Einstein equation, which follows from the 
action
\bea
\label{act1}
S = \int d^{5} x \sqrt{-\hat{G}}
\left\{ {1\over 16\pi G_5}\hat{R}+\Lambda \right\}\; .
\eea
Here $\hat{R}$ is the scalar curvature, $\Lambda$ is
the (positive) cosmological constant and $G_5$ denotes
the five--dimensional Newton constant.

The metric of the five--dimensional SdS black hole is
given by
\bea
\label{SdS}
ds^{2}_{5} &=& \hat{G}_{\mu\nu} dx^{\mu} dx^{\nu} \nn
&=& -\e^{2\rho} dt^2 + \e^{-2\rho} da^2
+ a^2 \sum_{ij}^{3}g_{ij}dx^{i}dx^{j} \ ,\nn
\e^{2\rho}&=&{1 \over a^{2} }\left( -\mu + {k\over 2}a^{2}
- {a^4 \over l^2} \right) ,
\eea
where
$l$ represents the curvature radius of the SdS bulk space 
and is related with  the cosmological constant 
$\Lambda={12 \over l^2}$. The three-dimensional metric, $g_{ij}$,
is the metric of an Einstein space with Ricci tensor given by
$r_{ij} =kg_{ij}$, where the constant $k=\{ -2, 0, +2 \}$
in our notations. The metric of the hypersurface with 
constant $a$ is then negatively-- curved, spatially flat,
or positively--curved depending on the sign of $k$. For SdS 
background, only $k=2$ is the static solution, but when 
$k=-2,0$, the coordinate $a$ plays a role of the (second) time 
coordinate and the solutions are time-evolving and 
cosmological ones with a big-bang singularity at $a=0$. 

The mass of the black hole is parametrized by a constant 
$\mu$, and $\mu$ can be expressed in terms of the horizon radius 
$a_H$:
\bea
\label{ab00}
\mu =a_{H}^{2} \left( -{a_{H}^{2} \over l^{2}}
+ {k \over 2} \right)\ .
\eea
The horizon radius $a_H$ is the solution of the equation
$\exp[2\rho (a_H)]=0$, which corresponds to (\ref{ab00}) 
\bea
\label{abrh1}
a_{H}^{2}={kl^{2} \over 4} \pm {1\over 2}
\sqrt{ {k^2 l^{4} \over 4} - 4 \mu l^{2} }
\eea
Note that, when $k=2$ SdS black hole has two  horizons 
$a_H$, that corresponds to the upper and lower signs in Eq.(\ref{abrh1}) 
(the cosmological and black hole horizons, 
respectively). Hereafter we denote black hole horizon by 
$a_{HB}$ and cosmological one by $a_{HC}$. When $k=0,-2$, there 
is no horizon since the right-hand side in (\ref{abrh1}) 
becomes imaginary or negative for positive $\mu$. Then in 
the following we consider mainly $k=2$ case. 

One can define  two Hawking temperatures corresponding 
to the two horizons:
\bea
\label{hwk}
T_{H} = \left| \left.{1\over 4\pi}{d \e^{2\rho}\over da}
\right|_{a=a_{BH},a_{CH}}\right|
= \left\{\begin{array}{lcl}
{1 \over 2\pi a_{BH}} - {a_{BH} \over \pi l^2} & \quad &
\mbox{for the black hole horizon} \\ 
-{1 \over 2\pi a_{CH}} + {a_{CH} \over \pi l^2} & \quad &
\mbox{for the cosmological horizon} \\ 
\end{array}\right. \ .
\eea
The Cardy-Verlinde (CV) formula \cite{EV} (see also \cite{cardy}) is
derived from the 
thermodynamical properties of the five--dimensional black hole.
So let us summarize the calculation of the thermodynamical
quantities like the free energy $F$, the entropy ${\cal S}$,
and the energy $E$ by following the method in \cite{no2}.  
After Wick-rotating the time variable
 $t\to i\tau$, the free energy $F$ can be obtained from the
action Eq.(\ref{act1}) as $F=-TS$. The classical solutions for
$\hat{R}$ and $\Lambda$ are given by $\hat{R}={20 \over l^2}$
and $\Lambda =-{12 \over 16\pi G_5 l^2}$. Then
the classical action  (\ref{act1}) takes the form
\bea
\label{act12}
S &=& {8 \over 16\pi G_5 l^2 } \int d^{5} x \sqrt{-\hat{G}} \; , \nn
&=& {W_{3} \over T} {8 \over 16\pi G_5 l^2 }
\int_{a_H}^{\infty} da \; a^3 \; ,
\eea
where $W_3$ is the volume of the unit three--sphere and
 $\tau$ has a period of ${1\over T}$. The expression
for $S$ contains the divergence coming from large $a$. In order
to subtract the divergence, we regularize $S$ (\ref{act12})
by cutting off the integral at a large radius $a_{\rm max}$
and subtracting the solution with $\mu =0$
in the same way as in \cite{no2}:
\bea
\label{13}
S = {W_{3} \over T}{8 \over 16\pi G_5 l^2}
\left\{ \int_{a_H}^{a_{\rm max}} da \; a^3 -\e^{\rho(a_{\rm max})
-\rho (a_{\rm max};\mu=0) } \int _{0}^{a_{\rm max}} da \; a^3 \right\} \; .
\eea
The factor $\e^{\rho(a_{\rm max})- \rho (a_{\rm max};\mu=0)}$
is chosen so that the proper length of the circle which
corresponds to the period ${1\over T}$ in the Euclidean time
at $a=a_{\rm max}$ coincides with each other in the two solutions.
Taking $a_{\rm max} \to \infty$, one finds
\be
F =\left\{\begin{array}{lcl}
-{8 W_{3} \over 16\pi G_5 l^2}\left( -{l^2 \mu \over 8}
-{a_{BH}^4\over 4} \right) &\quad & \mbox{for the black 
hole horizon}\ (a_H=a_{BH}) \\
{8 W_{3} \over 16\pi G_5 l^2}\left( -{l^2 \mu \over 8}
-{a_{CH}^4\over 4} \right) &\quad & \mbox{for the cosmological 
horizon}\ (a_H=a_{CH}) \\
\end{array}\right.\; .
\ee
One can rewrite $F$ by using Eq.(\ref{ab00}) as
\be
F =\left\{\begin{array}{lcl}
{W_{3} a_{BH}^2 \over 16\pi G_5}\left( {a_{BH}^2\over l^2} 
+ 1 \right) & \quad & \mbox{for black hole horizon} 
\ (a_H=a_{BH}) \\
-{W_{3} a_{CH}^2 \over 16\pi G_5}\left( {a_{CH}^2\over l^2} 
+ 1 \right) & \quad & \mbox{for cosmological horizon} 
\ (a_H=a_{CH}) \end{array}\right. \; .
\ee
The entropy ${\cal S}$
and the thermodynamical energy $E$ are
\bea
\label{ab3}
{\cal S} &=& -{dF\over dT_{H}}=-{dF\over da_{H}}{da_{H} \over dT_{H}} \nn
&=&{W_{3} a_{H}^3 \over 4 G_5} \quad \begin{array}{c}
\mbox{for both of black 
hole and cosmological horizons} \nn
(a_H=a_{BH}, a_{CH})\end{array} \; ,\\
\label{ab4}
E&=& F + T_{H} {\cal S} = \pm{3W_{3}\mu \over  16 \pi G_5 }
\quad \begin{array}{l}\mbox{$+$ for black hole horizon} \\
\mbox{and $-$ for cosmological one}
\end{array}\; .
\eea
Note that there can be two definitions of the temperature, 
the entropy and the energy, associated with two
horizons. From Appendix \ref{A2}, we find that the future
black hole horizon and the future cosmological horizon are
causally separated from each other. Then it is clear that any
particle, which exists between the black hole horizon and
the cosmological horizon, always may pass through only one of two future
horizons. The particle which crosses the black hole
horizon  observes the temperature and the entropy
associated with the black hole horizon but the other
particle which crosses the cosmological horizon 
observes the thermodynamical quantities associated with
the cosmological horizon.

In this paper, we are interested primarily in the
corrections to the entropy (\ref{ab3}) that
arise due to thermal fluctuations.
The leading--order correction
has been found for a generic thermodynamic system \cite{log11}.
The entropy is calculated in terms of a grand
canonical ensemble, where the corresponding
density of states, $\rho$, is determined
by performing an inverse Laplace transformation
of the partition function\footnote{The reader is referred to Refs.
\cite{log11,Log}
for details and related discussion in Refs.\cite{log6}}. The integral that
arises in this procedure
is then evaluated in an appropriate saddle--point
approximation. The correction to the entropy follows by assuming that
the scale, $\epsilon$, defined such that ${\cal{S}} \equiv
\ln (\epsilon \rho )$, varies in direct proportion
to the temperature, since this latter parameter is the only
parameter that provides a physical measure of scale in the
canonical ensemble.
The final result is then of the form \cite{Log}:
\begin{equation}
\label{gencorr}
{\cal{S}} = {\cal{S}}_0 -\frac{1}{2} \ln C_v + \ldots  ,
\end{equation}
where $C_v$ is the specific heat of the system evaluated at
constant volume and ${\cal S}_0$ represents the uncorrected entropy.
The derivation of (\ref{gencorr}) is given in Appendix \ref{A1}.
In the case of the SdS black hole, the entropy is given by
Eq. (\ref{ab3}). The specific heat of the black hole is determined
in terms of this entropy:
\bea
\label{sh}
C_{v} \equiv {dE \over dT_H} = 3 {2a_H^2-l^2 
\over 2a_H^2 + l^2}{\cal S}_0\ .
\eea
The above expression (\ref{sh}) is valid for both of 
the black hole and cosmological cases. 
For consistency, the condition $a_{H}^2 > l^2  /2$ should be 
satisfied to ensure that the specific heat is positive. In the 
limit $a_{H}^2 \gg l^2 /2$, $C_v
\approx 3 {\cal{S}}_0$, and this implies that \cite{Log}
\bea
\label{total}
{\cal S} = {\cal S}_0
-{1\over 2} \ln {\cal S}_0+ \cdots  \; .
\eea

Using the form of the logarithmic correction (\ref{total}) to 
the entropy, it is now possible to derive the corresponding 
corrections to  CV formula. We begin by recalling that
the four--dimensional energy, which can be derived from the 
FRW equation of motion for a brane propagating in an SdS bulk is given by
\bea
\label{e444}
E_{4} &=& \pm {3 W_{3} l \mu \over  16 \pi G_5 a}
\eea
($+$ corresponds to the black hole horizon and $-$ to the 
cosmological one) and is related to the five--dimensional 
energy (\ref{ab4}) of the bulk black hole such that 
$E_{4} =(l/a) E$ \cite{SV}. This implies that the temperature $T$, associated 
with the brane should differ from the Hawking temperature 
(\ref{hwk}) by a similar factor \cite{SV}:
\be
\label{abe22}
T={l \over a}T_H
= \left\{\begin{array}{lcl}
-{a_{BH} \over \pi a l} + {l \over 2\pi a a_{BH}} 
&\quad & \mbox{for the black hole horizon} \\
{a_{CH} \over \pi a l} - {l \over 2\pi a a_{CH}} 
&\quad & \mbox{for the cosmological horizon} \\
\end{array}\right.\ .
\ee

In determining the corrections to the entropy, a crucial physical 
quantity is the Casimir energy $E_C$ \cite{EV}, defined in terms
of the four--dimensional energy $E_4$, pressure $p$, volume 
$W=a^3 W_3$, temperature $T$ and entropy ${\cal{S}}$:
\be
\label{abEC1}
E_C=3\left( E_4 + pW - T{\cal S}\right)\ .
\ee
This quantity vanishes in case that the energy and entropy are 
purely extensive, but in general, this condition does not hold.
For the present discussion, the total entropy is assumed to be 
of the form (\ref{total}), where the uncorrected entropy 
${\cal{S}}_0$, corresponds to that associated with the black hole
in Eq. (\ref{ab3}) (due to the dS/CFT correspondence).
It then follows by employing (\ref{e444}) and (\ref{abe22})
that the Casimir energy (\ref{abEC1}) can be expressed  
in terms of the uncorrected entropy:
\be
\label{abEC2}
E_C= \pm \left({3 l a_H^2 W_3 \over 8 \pi G_5 a}
+{3\over 2}  T \ln {\cal S}_0\right) \ ,
\ee
where the direct dependence on the pressure has been
eliminated by assuming the relation $p=E_4/(3W_{3})$, which tells 
the 4d matter is the conformal one.

Moreover, in the limit where the logarithmic correction in Eq.
(\ref{abEC2})  is small, it can be shown, after substitution
of Eqs. (\ref{ab3}), (\ref{e444}) and (\ref{abEC2}), that the 
four--dimensional and Casimir energies are related to the 
uncorrected entropy by \cite{NOO}
\be
\label{abSS}
{4\pi a \over 3 \sqrt{2} }\sqrt{\left|
E_C\left( E_{4} - {1\over 2} E_C\right)\right|}
\sim \; {\cal S}_0 + \; {\pi a l \over 2 a_H^3} T
\left( {a_H^4 \over l^2} + a_H^2 \right) \ln {\cal S}_0 .
\ee
In the limit where the correction is small, the coefficient 
of the logarithmic term on the right--hand side of Eq.(\ref{abSS}) 
can be expressed in terms of the four--dimensional and Casimir 
energies through the relationship:
\be
\label{SN2}
{\pi a l \over 2 a_H^3} T \left( {a_H^4 \over l^2} + a_H^2 \right)
= {\left( 4 E_4 -E_C \right) \left( E_4 -E_C \right)
\over 2 \left( 2E_4 -E_C \right) E_C }  ,
\ee
where we have substituted  Eq. (\ref{abe22}) for the temperature and
have also employed the relation
\be
\label{SN1}
{E_4 - {1 \over 2}E_C \over E_C}= - {a_H^2 \over 2l^2}\ .
\ee
We may conclude, therefore, that in the limit where the 
logarithmic corrections are sub--dominant, Eq. (\ref{abSS}) can 
be rewritten to express the entropy in terms of the 
four--dimensional and Casimir energies (corrected Cardy-Verlinde
formula derived in \cite{lidsey, NOO} for AdS black hole):
\bea
\label{SN3}
{\cal S}_0 &=&
{4\pi a \over 3 \sqrt{2} }\sqrt{\left| E_C\left( E_{4}
- {1\over 2} E_C\right)\right|} \nn
&& -{\left( 4 E_4 -E_C \right) \left( E_4 -E_C \right)
\over 2 \left( 2E_4 -E_C \right) E_C }
\ln \left({4\pi a \over 3 \sqrt{2} }\sqrt{\left| E_C\left( E_{4}
 - {1\over 2} E_C\right)\right|}\right)
\eea
and, consequently, the total entropy Eq. (\ref{total})
to first order in the logarithmic term,
is given by \cite{lidsey, NOO}
\bea
\label{ttol}
{\cal S}
&\simeq&  {4\pi a \over 3 \sqrt{2}} \sqrt{\left| E_C\left( E_{4}
 - {1\over 2} E_C\right)\right|}\nn
&& - {E_4 \left( 4E_4 -3E_C \right) \over 2\left( 2E_4 -E_C \right) E_C }
\ln \left({4\pi a \over 3 \sqrt{2} }\sqrt{\left| E_C\left( E_{4}
- {1\over 2} E_C\right)\right|}\right)\; .
\eea
It then follows that the logarithmic corrections to CV formula
are given by the second term in the right hand side of 
Eq.(\ref{ttol}) and the magnitude of this correction can be 
deduced by taking the logarithm of the original CV formula. As 
we saw in above discussion these corrections
are caused by thermal fluctuations of the dS black hole.

If we consider the brane universe in the SdS black hole 
background, the four--dimensional FRW equation, which describes 
the motion of the brane universe, also receives corrections as 
a direct consequence of the logarithmic correction arising in 
Eq.(\ref{ttol}). In general, the Hubble parameter $H$ is 
related with the four--dimensional (Hubble) entropy (which is
identified with bulk black hole entropy, 
 see corresponding proof for AdS bulk in Ref. \cite{SV} and for dS bulk-in 
 \cite{SO}, For a brief review in the context of the 
brane world, see Appendix \ref{A3})
\bea
\label{HS}
H^2 = \left({2 G_4 \over W }\right)^2 {\cal S}^2\; ,
\eea
Here the effective four--dimensional Newton constant
$G_4$ is related to the five--dimensional Newton constant
$G_5$ by $G_4=2G_5/l$. The formula (\ref{HS}) is 
correct in both cases: when the brane crosses black hole horizon 
($a=a_{BH}$) and when the brane crosses the cosmological 
one ($a=a_{CH}$). As we will see in the next section, we extend 
the equation corresponding to (\ref{HS}) to the case for 
$a\neq a_{BH}$, $a_{CH}$. The extended equation (\ref{friedmannlog}) 
coincides with (\ref{HS}) in both cases $a=a_{BH}$ and $a=a_{CH}$.  
By substituting Eq. (\ref{ttol}) 
into Eq. (\ref{HS}), it can be shown by employing 
Eqs. (\ref{ab00}), (\ref{ab3}), (\ref{abe22}), (\ref{abEC2}), 
(\ref{SN3}) and (\ref{ttol}) that the four--dimensional FRW 
equation is 
\bea
\label{lnln}
H^2 &=& \left({2 G_4 \over W }\right)^2 \left[
\left( {4\pi a \over 3 \sqrt{2}} \right)^2
\left| E_C\left( E_{4} - {1\over 2} E_C\right)\right| \right.
-{4\pi a \over 3 \sqrt{2}}
{E_4 \left( 4E_4 -3E_C \right) \over \left( 2E_4 -E_C \right) E_C }\nn
&& \times \left.\sqrt{\left| E_C\left( E_{4}
 - {1\over 2} E_C\right)\right|}
\ln \left({4\pi a \over 3 \sqrt{2} }\sqrt{\left| E_C\left( E_{4}
 - {1\over 2} E_C\right)\right|}\right)\ \right] \nn
&=& {1 \over a_H^2} -{8\pi G_4 \over 3} \rho
-{2G_4 \over Wl} \ln {\cal S}_0\;.
\eea
Here the logarithmic corrections have been included
up to first--order in the logarithmic term,
the effective energy density
is defined by $\rho=|E_4|/W$ and
$W=a_H^3 W_3$ parametrizes the spatial volume of the
world--volume of the brane\footnote{
Since, at least, the brane receives the thermal radiation from the 
black hole, the thermal correction should change the dynamics of the 
brane from the leading order or zero temperature behavior. Since the 
detailed mechanism is not clear, here we naively assume that Eq.(\ref{HS}) is 
valid even if we include the thermal logarithmic correction. }.
In the limit where the scale factor $a$ of the brane
coincides with the horizon radius $a_H$ of the black hole,
the first and second terms on the right--hand--side of Eq. (\ref{lnln})
are identical to the FRW equation for the {\it space-like} brane
in SdS background whose signs of the terms
are the opposite to SAdS background \cite{SO, NOO}.
It is interesting that even if we did not assume the
space-like brane, the Hubble equation (\ref{lnln})
agrees with the case for space-like brane which
is the brane for the Wick-rotated version of standard 
FRW equation.

Then the logarithmic corrections for the FRW equation are
given by the third term on the right--hand side in terms of 
the uncorrected entropy (\ref{ab3}) of the black hole.

In the usual four-dimensional cosmology, the (first) FRW equation is given by
\bea
\label{FRWus}
H^2&=&{8\pi G \over 3}\rho - {1 \over a^2} \ ,\nn
\rho&=&\rho_m + {\Lambda \over 8\pi G}\ .
\eea
Here $\Lambda$ is a cosmological constant and $\rho_m$ corresponds to the
energy density of the matter. Typically in case that the matter is
radiation,
$\rho$ is proportional to ${1 \over a^4}$. Then Eq.(\ref{lnln}) tells
that, if we neglect the logarithmic correction, the obtained
energy density corresponds to the radiation and the
cosmological constant  should vanish. On the
other hand, by comparing (\ref{lnln}) and (\ref{FRWus}),
the logarithmic correction can be regarded as a small
effective cosmological constant by identifying
\be
\label{lnLmbd}
\Lambda_{\ln}=-{6G_4 \over Wl} \ln {\cal S}_0\ ,
\ee
although it depends on the size of the universe
since $W\propto a^3$. As $\rho$ behaves as $1 \over a^4$,
the $\Lambda_{\ln} \propto {1 \over a^3}$ varies slowly with the
expansion of the universe when $a$ is small. For large
$a$, where the approximation used here might not be valid,
$\Lambda_{\ln}$ becomes dominant if compare with the radiation.

Particularly, we consider the Nariai black hole which is
the most simple case. In this case, the second term
of (\ref{abrh1}) is zero, namely $a_{H}^2 ={l^2 \over 2}$.
Then black hole horizon coincides with cosmological horizon.

The Hubble equation (\ref{lnln}) for this case
looks
\bea
H^2 &=& {1  \over l^2}-{16 G_4 \over 2^{3/2} W_3 l^4 } \ln {\cal S}_0 \; ,\\
{\cal S}_{0} &=& {W_3 2^{3/2} l^2 \over 16 G_4 }\; .
\eea

In the Nariai limit where $\mu={l^2 \over 4}$, the expression (\ref{hwk}) 
for the Hawking temperature seems to vanish but this might not be true since 
$\e^{2\rho}=0$ in the region between the black hole and cosmological 
horizons, which tells that the coordinates $t$ and $a$ are degenerate 
or ill-defined in the region. Since the SdS solution is not asymptotically
flat, 
there is an ambiguity to rescale the time coordinate by a constant factor. 
We now introduce the following new coordinates $\tilde a$ and $\tilde t$:
\be
\label{rs1}
a^2= {l^2 \over 2}+ {\tilde a \over 2}
\sqrt{ l^{4}  - 4 \mu l^{2} } \ ,\quad
t={\tilde t \over \sqrt{ l^{4}  - 4 \mu l^{2} } }\ ,
\ee
Then there are horizons at $\tilde a=\pm 1$ and the metric has 
the following form:
\bea
\label{rs2}
ds^2 &=& - {\left(1 - \tilde a^2\right)d\tilde t^2 
\over 4l^2 \left({l^2 \over 2}+ {\tilde a \over 2}
\sqrt{ l^{4}  - 4 \mu l^{2} }\right)}
+ {l^2 d\tilde a^2 \over 4\left(1 - \tilde a^2\right)} \nn
&& + \left( {l^2 \over 2}+ {\tilde a \over 2}
\sqrt{ l^{4}  - 4 \mu l^{2} } \right)
\sum_{ij}^{3}g_{ij}dx^{i}dx^{j} \ ,
\eea
and the Hawking temperature $T_H$ is also rescaled as
\be
\label{rs3}
T_H\to \tilde T_H \equiv {T_H \over \sqrt{ l^{4}  - 4 \mu l^{2}}}
= {1 \over 2\pi l^2 a_H}\ .
\ee
Then in the Nariai limit, the metric has the following form:
\be
\label{rs4}
ds^2 = - {\left(1 - \tilde a^2\right) \over 2l^4 } d\tilde t^2
+ {l^2 d\tilde a^2 \over 4\left(1 - \tilde a^2\right)}
+ {l^2 \over 2} \sum_{ij}^{3}g_{ij}dx^{i}dx^{j} \ ,
\ee
and the rescaled Hawking temperature is finite
\be
\label{rs5}
T_H={1 \over \pi l^3\sqrt{2}}\ .
\ee
If we further rewrite the coordinate $\tilde a$ as $\tilde a=-\cos\theta$, 
we obtain 
\be
\label{rs6}
ds^2 = - {\sin^2 \theta \over 2l^4 } d\tilde t^2
+ {l^2 d\theta^2 \over 4} + {l^2 \over 2} \sum_{ij}^{3}g_{ij}dx^{i}dx^{j} \ ,
\ee
which might be a standard form of the metric in the Nariai space.

\section{Qualitative Dynamics of the Brane Cosmology \label{Sec2}}

In this section, we investigate the
asymptotic behavior of the FRW brane cosmology
when the logarithmic corrections to the CV formula
are included. Formally, the FRW equation
(\ref{lnln}) holds precisely at the instant when the brane crosses
black hole and cosmological horizons. Here we extend the analysis
to consider an arbitrary scale factor $a$ where the world--volume
of the brane is given by the line--element
$ds^2_4 = d\tau^2 +a^2(\tau)g_{ij}dx^{i}dx^{j}$. 
Thus, around each horizon
we assume the FRW equation as follows:
\bea
\label{friedmannlog0}
H^2 = {1 \over a^2} -{8\pi G_4 \over 3} \rho
-{2 G_4 \over W l} \ln {\cal S}_0\;,
\eea
where $W=a^3 W_3$ and ${\cal S}_0 =W_3 a^3 /(4G_5)$.
This equation differs by several signs from the corresponding FRW brane
equation with log-corrections obtained in Ref.\cite{lidsey} where bulk is
 AdS black hole.
On the black hole and cosmological horizon
Eq.(\ref{friedmannlog0}) agrees with Eq.(\ref{lnln}).
We also extend the result of the previous section to the 
general $k$:
\bea
\label{friedmannlog}
H^2 = {k \over 2 a^2} -{8\pi G_4 \over 3} \rho
-{2 G_4 \over W l} \ln {\cal S}_0\;,
\eea
Eq.(\ref{friedmannlog}) can be rewritten in such a way
that it represents the conservation of energy of a point particle
moving in a one--dimensional effective potential, $V(a)$:
\bea
\left( {da \over d\tau }\right)^2 &=& {k \over 2} -V(a) \\
\label{effec}
V(a) &\equiv& {8\pi G_4 \over 3}a^2 \rho + {2G_4 a^2 \over Wl}
\ln {\cal S}_0 \nn
&=& {\mu \over a^2} + {2G_4 \over W_3 l a}\ln \left({W_3 a^3 \over 
2l G_4}\right)\ ,
\eea
where, in this interpretation, the variable $a$ represents the
position of the particle.
Since $\rho \propto a^{-4}$, the first term in
the effective potential (\ref{effec}) redshifts as
$a^{-2}$ as the brane moves away from the black hole horizon.
This term is often referred to as the `dark radiation' term.

\begin{figure}[htbp]
\begin{center}
\unitlength=0.47mm
\begin{picture}(160,110)
\thicklines
\put(30,93){$V(a)$}
\put(152,38){$a$}
\put(40,40){\vector(1,0){110}}
\put(40,10){\vector(0,1){80}}
\qbezier[400](42,90)(45,60)(60,50)
\qbezier[400](60,50)(72,43)(150,41)
\put(3,48){${k \over 2}=1$}
\put(3,38){${k \over 2}=0$}
\put(3,28){${k \over 2}=-1$}
\put(50,33){$a=a_{\rm min}$}
\thinlines
\put(40,50){\line(1,0){110}}
\put(40,30){\line(1,0){110}}
\put(60,40){\line(0,1){10}}

\thinlines

\qbezier[400](42,90)(45,65)(60,55)
\qbezier[400](60,55)(72,45)(150,41)

\end{picture}
\end{center}
\caption{\label{Fig1}
The effective potential for
the FRW brane Universe in SdS bulk. For  $k=2$,
the spherical brane starts at $a=\infty$ and
reaches its minimum size at $a=a_{\rm min}$ and then it
re-expands. For Nariai black hole, the effective
potential is larger than that in the other SdS cases 
as illustrated by thin line. }
\end{figure}
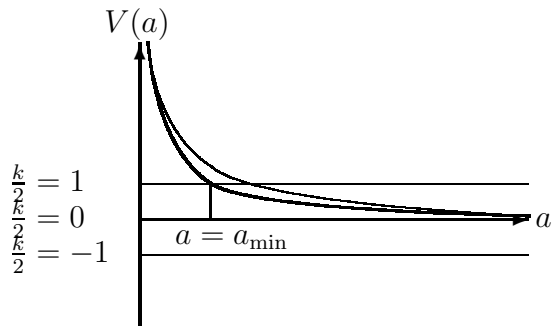

To proceed in the analogy with \cite{lidsey}, let us briefly
recall the behavior of the standard FRW cosmology, whose effective
potential includes only the first term on the right--hand side
of Eq. (\ref{effec}). The behavior of this
potential is illustrated in Figure \ref{Fig1}. The brane
exists in the regions where the line $k/2 \geq V(a)$
(so that $H^2 > 0$). Then, we only have the case of $k=2$
which is the spherical brane. The spherical brane
starts from $a=\infty$ and reaches its minimum size
at $a=a_{\rm min}=\sqrt{\mu}$ and then it re-expands.

\begin{figure}[htbp]
\begin{center}
\unitlength=0.47mm
\begin{picture}(160,110)
\thicklines
\put(30,93){$V(a)$}
\put(152,38){$a$}
\put(40,40){\vector(1,0){110}}
\put(40,10){\vector(0,1){80}}
\qbezier[400](42,90)(45,20)(60,25)
\qbezier[400](60,25)(65,30)(70,45)
\qbezier[400](70,45)(80,65)(110,50)
\qbezier[400](110,50)(120,45)(150,43)
\put(3,48){${k \over 2}=1$}
\put(3,38){${k \over 2}=0$}
\put(3,28){${k \over 2}=-1$}

\thinlines

\qbezier[400](42,90)(45,25)(60,33)
\qbezier[400](60,33)(65,35)(70,43)
\qbezier[400](70,43)(80,50)(110,45)
\qbezier[400](110,45)(120,42)(150,41)

\put(40,50){\line(1,0){110}}
\put(40,30){\line(1,0){110}}
\end{picture}
\end{center}
\caption{\label{Fig2}
The effective potential for
the FRW Universe in SdS bulk
when  logarithmic corrections are included.
There are several cases which depend on
the parameters $G_4,W_3,l$.}
\end{figure}
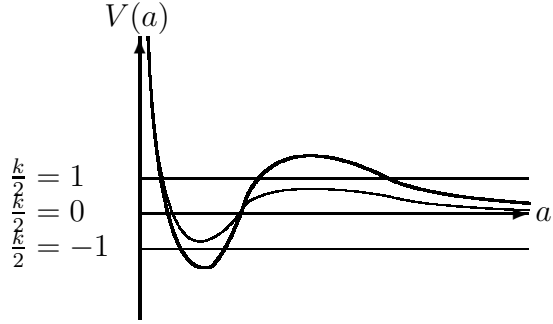

 From Eq.(\ref{e444}) the energy density $\rho=E_4/W$ looks like
${3\mu \over 8\pi G_4 a^4}$, then the first term in
Eq.(\ref{effec}) is rewritten as $\mu/a^2$:
\bea
\label{effec2}
V(a) &\equiv& {\mu \over a^2} + {2G_4 \over W_3 l a}
\ln {\cal S}_0\; .
\eea
For the Nariai black hole, the mass $\mu$ takes
the particular value $\mu={k^2 l^2 \over 16}$
which is the largest mass for the SdS black hole,
since the inside of square root in Eq.(\ref{abrh1})
must be positive. Then the behavior of the
potential for Nariai BH is bigger than in
SdS case as illustrated by thin line in Figure \ref{Fig2} 
so that its minimum size $a_{\rm min}$ is bigger
than the minimum size of SdS.

Next, one considers the behavior of
the effective potential with logarithmic corrections. From 
Eq.(\ref{effec2}), there are several cases which depend on
the parameters $G_4,W_3,l,\mu$. If the coefficient ${2G_4 \over W_3 l}$ 
of second term in Eq.(\ref{effec2}) is equal to or less than $\sqrt{\mu}$,
the behavior of the potential is not so changed from Figure \ref{Fig1}.
But when the coefficient ${2G_4 \over W_3 l}$ of the second term in 
Eq.(\ref{effec2}) is large compared with $\sqrt{\mu}$,
the behavior of the potential changes from thin line to
thick line as illustrated in Figure \ref{Fig2}.

For Nariai black hole, the ratio of $\sqrt{\mu}$ and
the coefficient of the second term in Eq.(\ref{effec2}) is smaller
than that of SdS case since $\mu$ is
always bigger than the mass of SdS black hole.
Then the behavior of the effective potential
is similar to Figure \ref{Fig2} but smaller than
that of SdS case.

As an explicit example one can take five-dimensional deSitter background
instead of SdS background. The dS metric is given by
\bea
\label{dS1}
ds^{2}_{5} &=& -\e^{2\rho} dt^2 + \e^{-2\rho} da^2
+ a^2 \sum_{ij}^{3}g_{ij}dx^{i}dx^{j} \ ,\nn
\e^{2\rho}&=& 1 - {a^2 \over l^2} \; ,
\eea
which is the massless case $\mu=0,\; k=2$ in Eq.(\ref{dS1}).
Then, horizon radius looks like $a_H = l$.
From Eq.(\ref{lnln}), the FRW equation for dS case also takes simple form
as
\bea
\label{lnln2}
H^2 ={1 \over a_H^2}-{2G_4 \over Wl} \ln {\cal S}_0\;,
\eea
where ${\cal S}_0 = {W_3 a_H^3 \over  4 G_5}$.

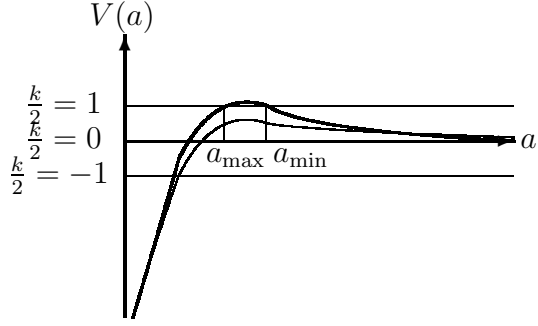
\begin{figure}[htbp]
\begin{center}
\unitlength=0.47mm
\begin{picture}(160,110)
\thicklines
\put(30,93){$V(a)$}
\put(152,58){$a$}
\put(40,60){\vector(1,0){110}}
\put(40,10){\vector(0,1){80}}
\qbezier[400](42,10)(48,30)(55,55)
\qbezier[400](55,55)(65,75)(80,70)
\qbezier[400](80,70)(90,63)(150,60)

\put(12,68){${k \over 2}=1$}
\put(12,58){${k \over 2}=0$}
\put(7,48){${k \over 2}=-1$}
\put(63,54){$a_{\rm max}$}
\put(82,54){$a_{\rm min}$}
\thinlines

\qbezier[400](42,10)(48,30)(55,50)
\qbezier[400](55,50)(65,70)(80,65)
\qbezier[400](80,65)(90,63)(150,61)

\put(40,70){\line(1,0){110}}
\put(40,50){\line(1,0){110}}
\put(68,60){\line(0,1){10}}
\put(80,60){\line(0,1){10}}
\end{picture}
\end{center}
\caption{\label{Fig3}
The behavior of the effective potential for FRW 
Universe in deSitter bulk with logarithmic corrections. 
There are two types of behavior for spherical brane $k=2$.
For the case of thick line, the brane starts from $a=0$
and reaches its maximal size $a=a_{\rm max}$ and then it re-collapses,
or the brane starts from $a=\infty$ and reaches its minimum size
at $a=a_{\rm min}$ then it re-expands.
For the case of thin line, the brane starts from $a=0$ and
expands to infinity.}
\end{figure}

One can extend the FRW equation to general $a$ and 
$k$:
\bea
\label{lnln3}
H^2 ={k \over 2a^2}-{2G_4 \over Wl} \ln {\cal S}_0\; .
\eea
Here ${\cal S}_0 = {W_3 a^3 \over  4 G_5}$, $W = a^3 W_3$, again.
This equation defines the effective potential $V(a)$ as
\bea
V(a) = {2G_4 \over a W_3 l} \ln {\cal S}_0 \; .
\eea
The behavior for the effective potential for dS bulk
is illustrated in Figure \ref{Fig3}. When $k=0,-2$, the brane starts 
from $a=0$ and reaches its maximal size $a_{\rm max}$ and then it
re-collapses. Note that the behavior of the effective potential
with logarithmic correction for FRW universe in deSitter bulk differs
from the one in SdS bulk (Figure \ref{Fig2}).
If the maximum of the effective potential $V_{\rm max}$ is larger 
then 1, there are two solutions for $k=2$ case. In one case, the 
brane started at $a=0$ reaches its maximum and shrinks. In another 
case, the brane started at $a=\infty$ shrinks and reaches its minimum 
and reexpands, which is the bounce universe case. 

When there are no logarithmic corrections, Eq.(\ref{lnln3}) has 
a simple form:
\be
\label{lnlnln1}
\dot a^2={k \over 2}\ ,
\ee
Then when $k=-2$, there is no solution, when $k=0$, the brane is static 
($\dot a=0$). When $k=2$, the solution is given by
\be
\label{lnlnln2}
a=|\tau|\ .
\ee
The solution has a singularity at $\tau=0$. The second, logarithmic 
correction term in (\ref{lnln3}) can be neglected compared with 
the first term for large $a$. Then the behavior of the brane with $k=2$ 
when $a$ is large is given by (\ref{lnlnln2}) even if we include 
the logarithmic correction. When $a$ is small, the logarithmic 
correction becomes dominant. Then (\ref{lnln3}) can be approximated as
\be
\label{lnlnln3}
\dot a^2 \sim -{2G_4 \over aW_3l} \ln {W_3 a^3 \over  4 G_5}\; ,
\ee
Then $a$ behaves as $a\sim |\tau|^{2 \over 3}$ (by approximating 
$\ln a$ to be constant, compared with ${1 \over a}$).

Thus, we demonstrated that the evolution of  spherical brane 
which could correspond to our observable early Universe depends 
explicitly from 
the choice of bulk (dS, SdS or Nariai space) and from
the inclusion (or not) of log-corrections.   

\section{Logarithmic corrections for four-dimensional FRW cosmology}

In this section, we forget for the moment about the braneworld and 
discuss the role of logarithmic corrections to usual 4d cosmology 
and to 4d CV formula (see \cite{EV} and for CV formula in 4d dS space,
see\cite{abdalla}).
One starts from 
the Einstein gravity with positive cosmological 
constant $\Lambda_4>0$. Then the standard FRW equation 
has the following form:
\be
\label{AA1}
H^2 = {8\pi G_4 \over 3}\rho_m - {1 \over a^2} 
+ {1 \over l^2}\ .
\ee
Here $\rho_m$ is the energy density of the matter  and the 
length parameter $l$ is given by $\Lambda_4={3 \over l^2}$. 
We also consider only $k=2$ case. If the matter energy can be 
neglected as $\rho_m\ll {3 \over 8\pi G_4 l^2}$, the spacetime becomes 
deSitter space (in the static coordinates)
\be
\label{AA2}
ds^2 = - \e^{2\rho}dt^2 + \e^{-2\rho}da^2 
+ a^2 d\Omega_2^2\ , \quad \e^{2\rho}
\equiv 1 - {a^2 \over l^2}\ .
\ee
Here $d\Omega_2^2$ is the metric of the two-sphere. Then 
the cosmological horizon is given by $a=a_H\equiv l$ and 
the Hawking temperature $T_H$ is defined as
\be
\label{AA3}
T_H=\left|{1 \over 4\pi} \left.{d\e^{2\rho} \over da}\right|_{a=l}
\right|={1 \over 2\pi l}\ .
\ee
Then the entropy is found to be
\be
\label{AA4}
S_0={\pi l^2 \over G_4}\ .
\ee
The expression for the Casimir energy $E_C$ 
\be 
\label{AA5}
E_C=3\left(E+pV - TS\right)\ ,
\ee
suggests that the logarithmic correction to 
the entropy 
\be
\label{AA6}
S_0\to S_0 - {1 \over 2}\ln S_0
\ee
may shift the energy by
\be
\label{AA6b}
\delta E = - {T \over 2}\ln S_0 = -{1 \over 4\pi l}\ln 
{\pi l^2 \over G_4}\ .
\ee
This suggests the following modification of FRW equation (\ref{AA1}) 
\be
\label{AA7}
H^2 = {8\pi G_4 \over 3}{\delta E \over V} - {1 \over a^2} 
+ {1 \over l^2}
= - {G_4 \over 3\pi^2 l a^3}\ln {\pi l^2 \over G_4} - {1 \over a^2} 
+ {1 \over l^2}\ .
\ee
Here $V=2\pi^2 a^3$ is the volume of the three-sphere. 
The correction to the energy effectively shifts 
the cosmological constant and might be dominant for 
small $a$. When ${\pi l^2 \over G_4}>1$, the effective 
cosmological constant decreases due to the correction 
and when ${\pi l^2 \over G_4}>0$, it increases. 
The $a^{-3}$ behavior of ${\delta E \over V_3}$ tells 
the correction part of the energy behaves as the (effective) dust where the 
pressure  vanish $p=0$. 

If we assume $E=\delta E$ (in the absence of matter), by using (\ref{AA5}) 
one obtains 
an expression for the Casimir energy:
\be
\label{AA8}
E_C= - {24\pi^2 l \over G_4}\ .
\ee
The expression (\ref{AA8}) is not changed by the 
logarithmic correction. 
Then,  using (\ref{AA4}) one gets 
\be
\label{AA9}
S_0={l \over 24\pi}\left|E_C\right|\ ,
\ee
or  using (\ref{AA6}), we may obtain
\be
\label{AA10}
{\cal S}\equiv S_0 - {1 \over 2}\ln S_0 
= {l \over 24\pi} \left|E_C\right| - {1 \over 2}\ln\left(
{l \over 24\pi}
\left|E_C\right|\right)\ ,
\ee
which corresponds to the Cardy-Verlinde formula in the situation under
consideration. 
Even if the logarithmic correction is not included, 
the formula is rather different from the usual one:
\be
\label{AA11}
{\cal S}={l \over 24\pi}\left|E_C\right|\ .
\ee
We may compare the expression (\ref{AA10}) with the 
Cardy-Verlinde formula with the logarithmic correction  (\ref{ttol}), 
which has been obtained in the braneworld context.  
By putting $E_4=0$ in (\ref{ttol}), the logarithmic correction vanishes 
and we obtain
\be
\label{ttolB}
{\cal S}\simeq  {2\pi a \over 3} \left| E_C\right|\; ,
\ee
which is similar to (\ref{AA11}), rather than (\ref{AA10}), and 
identical with (\ref{AA11}) if we put $a={l \over 16\pi}$.   

Thus, we found that logarithmic correction to the entropy may lead to inducing
 of small effective cosmological constant in FRW equation. Eventually, 
this may have some cosmological applications.

\section{Discussion}

In the present paper we discussed the role of logarithmic corrections 
which appear in SdS black hole entropy to FRW brane cosmology.
The relation between black hole entropy and Hubble parameter is
controlled by dS/CFT correspondence.
These, next-to-leading corrections in FRW equation may be interpreted 
as small effective cosmological constant which qualitatively changes the
evolution of spherical brane. The examples of the spherical brane evolution 
are presented without (or with) logarithmic terms and for different bulk:
dS, SdS or Nariai space. Eventually, if our brane FRW Universe is embedded 
into SdS bulk (or AdS black hole \cite{lidsey}), these next-to-leading terms 
may be very important in cosmology as we explicitly demonstrated.
 
Let us now comment their role in the entropy bounds estimations. 
If we define the four-dimensional Hubble, Bekenstein-Hawking and Bekenstein
entropies by \cite{EV}
\be
\label{dsc1}
{\cal S}_H\equiv {HW \over 2G_4}\ ,\quad {\cal S}_{BH}\equiv 
{W \over 4G_4 a_H}\ , \quad {\cal S}_B\equiv {2\pi a \over 3}\rho W\ ,
\ee
we can rewrite (\ref{lnln}) as 
\be
\label{dsc2}
{\cal S}_H^2 = \left({\cal S}_{BH} - {\cal S}_B
\right)^2 - {\cal S}_B^2 - {\cal S}_C\ln {\cal S}_C\ .
\ee
Here ${\cal S}_C$ is defined, in a similar way to ${\cal S}_{BH}$, by
\be 
\label{dsc3}
{\cal S}_C\equiv {W \over 2G_4 l}={W \over 4G_5}\ ,
\ee
which may give a lower limit of ${\cal S}_{BH}$ since $a_H\leq l$ and 
we have $a_H=l$ for cosmological horizon in pure deSitter space 
($\mu=0$). Eq.(\ref{dsc3}) tells that ${\cal S}_C$ is the entropy 
of the 5d black hole, whose horizon area is equal to the space-like 
volume of the brane. 
We also note that if we conjecture the redefined Bekenstein-Hawking entropy 
as
\be
\label{dsc4}
{\cal S}_{BH}\to \hat{\cal S}_{BH}={\cal S}_{BH} + {{\cal S}_C 
\ln {\cal S}_C \over 2\left({\cal S}_{BH} - {\cal S}_B\right)}\ ,
\ee 
Eq.(\ref{lnln}) or (\ref{dsc2}) can be rewritten as
\be
\label{dsc2b}
{\cal S}_H^2 = \left(\hat{\cal S}_{BH} - {\cal S}_B
\right)^2 - {\cal S}_B^2 \ .
\ee
This looks as standard CV formula which defines the entropy bounds 
in expanding Universe but the interpretation of Bekenstein-Hawking entropy 
is changed.

Finally, one can note that similar considerations  are applied in the study 
of FRW or anisotropic brane cosmology with logarithmic corrections 
for another types of dS bulk black holes.

\section*{Acknowledgments}

We thank E. Abdalla for useful discussion.
The research by S.N. is supported in part by the Ministry of Education,
Science, Sports and Culture of Japan under the grant n. 13135208.
The  research by S.O. is supported in part by the Japan Society
for the Promotion of Science under the Postdoctoral Research Program.

\appendix
\addcontentsline{toc}{part}{Appendix}
\part*{Appendix}
\section{Logarithmic corrections to the entropy\label{A1}}

In this section, we review briefly the calculation of the 
log correction to the entropy.
First, let us recall the expression for the partition function in
the grand canonical ensemble 
by
\bea
\label{bun}
Z(\beta) = \int \e^{-\beta E} \rho(E) dE\; .
\eea
Here $\rho (E)$ is the density of states,
$\beta$ is the inverse temperature,
$\beta = {1\over T}$, and we set $k_{B}=1$,
so that the temperature has the dimension
of energy. From the classical thermodynamical
relation between free energy $F$, energy $E$
and entropy $S$:
\bea
F=E-TS\; , \quad F = -T \ln Z\; ,
\eea
the partition function Eq.(\ref{bun})
can be written as follows:
\bea
\label{kei}
\e^{-\beta F} = \int dE \; L\e^{-\beta E + S (E)} \; ,
\eea
where $\rho(E)=L\e^{S(E)}$. The parameter $L$ has the dimension of the 
length and can be determined by the details of the system under
consideration. 

The function $-\beta E + S (E)$ can be expanded around
the energy of thermal equilibrium point $E_0$ as
\be
 -\beta E + S (E) = -\beta E_0 + S (E_0)
+{1\over 2}\beta^{2} B(E_0)(E-E_0)^2
+{\cal O}\left((E-E_0)^3\right) \; .
\ee
Here the coefficient $B(E_0)$ is the dimensionless
constant related with $E_0$. Of course, $1 \over B(E_0)$
is nothing but the square of the standard deviation.
Then Eq.(\ref{kei}) can be calculated by Gaussian integral
up to second order on $(E-E_0)^{2}$ as
\bea
\e^{-\beta F} &=& \sqrt{\pi L^2\over \beta^2 B(E_0)} \;
\e^{-\beta E_0 + S (E_0)} \; ,\nn
&=& \e^{-\beta E_0 + S (E_0)+{1\over 2}\ln
{\pi L^2 \over \beta^2 B(E_0)} } \; ,
\eea
which leads to the relation
\bea
\label{FF}
-\beta F = -\beta E_0 + S (E_0)+{1\over 2}\ln {C(E_0) L^2\over \beta^2} \;
\eea
where $C(E_0)={\pi \over B(E_0)}\; $.
Since the specific heat is given by
\be
\label{SH}
\left.{\partial \left< E \right> \over \partial T}\right|_V
=\beta^2 \left(\left< E^2 \right> - \left< E \right>^2
\right) = {1 \over B(E_0)}
\ee
$C(E)\over \pi$ can be regarded as the specific heat.
If we assume the second and third terms in the right hand side of
Eq.(\ref{FF}) as uncorrected entropy $S_0$,
\bea
S_0 =  S (E_0)+{1\over 2}\ln {C(E_0) L^2 \over \beta^2}\; ,
\eea
the similar equation to (\ref{total}) is obtained:
\bea
S(E_0) =  S_0 - {1\over 2}\ln {C(E_0) L^2\over \beta^2}\; .
\eea
Therefore one can realize that the entropy $S(E_0)$ always has
the logarithmic correction from the classical thermodynamical 
considerations. 

If one can choose $L=\beta$, Eq.(\ref{gencorr}) is reproduced but this
might not 
be justified since $L$ might be determined independently from the 
temperature. Then there could be extra $\beta$ or temperature dependence 
inside the logarithmic function. In the case that we are considering in this 
paper, the temperature $T$ is determined by the radius of the horizon,
then $T$ might not depend on the radius $a$ of the brane universe or
$T$ might scale as $l/a$ as in Eq.(\ref{abe22}). Even in the latter case, 
only the power of $a$ inside the logarithmic term changes from 
$a^3$ to $a$, which changes the coefficient in front of the 
logarithmic term but the qualitative structure is be 
changed. 

For Schwarzschild-deSitter spacetime, there are two kinds of 
temperatures corresponding to two horizons, black hole horizon 
and cosmological one. The system is not thermodynamically stable. 
However, the system should be adiabatic since one 
can define the temperature in the vicinity of each horizon. 
Furthermore,  future black hole and cosmological horizons are 
separated from each other as we will see in the next Appendix 
\ref{A2}. Then we may discuss the thermodynamics near the 
horizon. 

\section{Penrose diagram for Schwarzschild-deSitter black hole.\label{A2}}

The Penrose diagram for Schwarzschild-deSitter black hole is given in 
Figure \ref{Fig4}. We can find the future black hole horizon is 
causally separated from the cosmological one. Then any particle 
in a region between the black hole and cosmological horizons will 
cross one and only one of the future horizons. Then such a particle 
 observes the energy (the entropy, etc.) associated with 
the horizon that the particle crosses. 
\unitlength=0.5mm

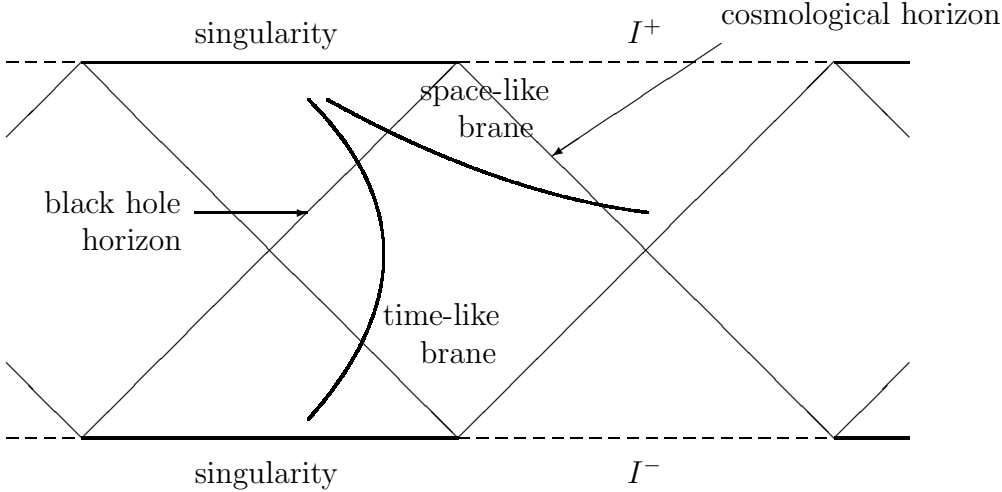
\begin{figure}
\begin{picture}(300,150)

\thinlines

\put(30,20){\line(-1,1){20}}
\put(30,20){\line(1,1){100}}
\put(30,120){\line(-1,-1){20}}
\put(30,120){\line(1,-1){100}}
\put(130,20){\line(1,1){100}}
\put(130,120){\line(1,-1){100}}
\put(230,20){\line(1,1){20}}
\put(230,120){\line(1,-1){20}}

\multiput(10,20)(5,0){4}{\line(1,0){3}}
\multiput(10,120)(5,0){4}{\line(1,0){3}}
\multiput(130,20)(5,0){20}{\line(1,0){3}}
\multiput(130,120)(5,0){20}{\line(1,0){3}}
\put(60,80){\vector(1,0){30}}
\put(200,125){\vector(-3,-2){45}}

\thicklines
\put(30,20){\line(1,0){100}}
\put(30,120){\line(1,0){100}}
\put(230,20){\line(1,0){20}}
\put(230,120){\line(1,0){20}}

\qbezier[600](90,110)(130,70)(90,25)
\qbezier[600](95,110)(140,85)(180,80)

\put(20,80){black hole}
\put(30,70){horizon}
\put(200,130){cosmological horizon}
\put(120,110){space-like}
\put(130,100){brane}
\put(110,50){time-like}
\put(120,40){brane}
\put(60,125){singularity}
\put(60,8){singularity}
\put(175,125){$I^+$}
\put(175,8){$I^-$}

\end{picture}
\caption{\label{Fig4} The time-like and space-like branes 
in the Penrose diagram of the Schwarzschild-deSitter spacetime.}

\end{figure}

\section{A brief review of the Cardy-Verlinde Formula in the context of the 
brane world \label{A3}}

In this appendix, we briefly explain how the Cardy-Verlinde formula 
can be understood in the context of the brane world in the SAdS bulk. 
For the SdS bulk case, 
see \cite{SO}. Here we do not include the logarithmic corrections. 

We start with the Minkowski signature action $S$ which is 
the sum of the Einstein-Hilbert action $S_{\rm EH}$ with the 
cosmological term, the Gibbons-Hawking surface term $S_{\rm GH}$, and
the surface counter term $S_1$: 
\bea
\label{Stotal}
S&=&S_{\rm EH} + S_{\rm GH} + 2 S_1\ , \\
\label{SEHi}
S_{\rm EH}&=&{1 \over 16\pi G}\int d^5 x \sqrt{-g_{(5)}}\left(R_{(5)} 
 + {12 \over l^2} \right)\ , \\
\label{GHi}
S_{\rm GH}&=&{1 \over 8\pi G}\int d^4 x \sqrt{-g_{(4)}}\nabla_\mu n^\mu, \\
\label{S1}
S_1&=& -{6 \over 16\pi G l}\int d^4 x \sqrt{-g_{(4)}}\ .
\eea 
Here the quantities in the  5 dimensional bulk spacetime are 
specified by the suffices $_{(5)}$ and those in the boundary 4 
dimensional spacetime are specified by $_{(4)}$. 
In (\ref{GHi}), $n^\mu$ is the unit vector normal to the 
boundary. The Gibbons-Hawking term $S_{\rm GH}$ is necessary 
in order to make the variational method well-defined when there is 
boundary in the spacetime. In (\ref{S1}), the coefficient of 
$S_1$ is determined from AdS/CFT. The factor 2 in front of $S_1$ 
is coming from that we have two bulk regions which 
are connected with each other by the brane. 
Then on the brane, we have the following equation:
\be
\label{eq2b}
0=A_{,z} - {1 \over l}\ .
\ee
This equation is derived from the condition that the variation 
of the action on the brane, or the boundary of the bulk spacetime, 
vanishes under the variation over $A$. The first term 
proportional to $A_{,z}$ expresses the bulk gravity force acting 
on the brane and the term proportional to ${1 \over l}$ comes 
from the brane tension. 
In (\ref{eq2b}), one uses the form of the metric as 
\be
\label{metric1}
ds^2=dz^2 + \e^{2A(z,\tau)}\tilde g_{\mu\nu}dx^\mu dx^\nu\ ,
\quad \tilde g_{\mu\nu}dx^\mu dx^\nu\equiv l^2\left(-d \tau^2 
+ d\Omega^2_3\right)\ .
\ee
Here $d\Omega^2_3$ corresponds to the metric of 3 dimensional 
unit sphere. 

As a bulk space, we consider 5d AdS-Schwarzschild black hole spacetime, 
whose metric is given by,
\be
\label{AdSS}
ds_{\rm AdS-S}^2 = {1 \over h(a)}da^2 - h(a)dt^2 
+ a^2 d\Omega_3^2 \ ,\ \ 
h(a)= {a^2 \over l^2} + 1 - {16\pi GM \over 3 V_3 a^2}\ .
\ee
Here $V_3$ is the volume of the unit 3 sphere.  
If one chooses new coordinates $(z,\tau)$ by
\bea
\label{cc1}
&& {\e^{2A} \over h(a)}A_{,z}^2 - h(a) t_{,z}^2 = 1 \ ,
\quad {\e^{2A} \over h(a)}A_{,z}A_{,\tau} - h(a)t_{,z} t_{,\tau}
= 0 \nn
&& {\e^{2A} \over h(a)}A_{,\tau}^2 - h(a) t_{,\tau}^2 
= -\e^{2A}\ .
\eea
the metric takes the warped form (\ref{metric1}). Here $a=l\e^A$.
In general we might be unable to  rewrite globally the metric in 
(\ref{AdSS}) in the form of (\ref{metric1}). Nevertheless, it can be done 
in the neighbourhood of the brane, what is necessary here. 
Further choosing a coordinate $\tilde t$ by 
$d\tilde t = l\e^A d\tau$, the metric on the brane takes FRW form: 
\be
\label{e3}
\e^{2A}\tilde g_{\mu\nu}dx^\mu dx^\nu= -d \tilde t^2  
+ l^2\e^{2A} d\Omega^2_3\ .
\ee
By solving Eqs.(\ref{cc1}), we have
\be
\label{e4}
H^2 = A_{,z}^2 - h\e^{-2A}= A_{,z}^2 - {1 \over l^2}
 - {1 \over a^2} + {16\pi GM \over 3 V_3 a^4}\ .
\ee
Here the Hubble constant $H$ is introduced:
$H={dA \over d\tilde t}$. 
By using (\ref{eq2b}), we find
\be
\label{e7}
H^2 = - {1 \over a^2} + {16\pi GM \over 3 V_3 a^4} \ .
\ee 
Further by differentiating Eq.(\ref{e7}) with respect to 
$\tilde t$, we obtain
\be
\label{e9}
H_{,\tilde t} =  {1 \over a^2} - {32\pi GM \over 3 V_3 a^4} \ .
\ee
One can rewrite the above equations (\ref{e7}) and (\ref{e9}) 
in the form of the standard FRW equations: 
\bea
\label{e10}
&& H^2 = - {1 \over a^2} 
+ {8\pi G_4 \rho \over 3} \ , \quad \rho= {Ml \over V_3 a^4}\ ,\\
\label{e11}
&& H_{,\tilde t} =  {1 \over a^2} - 4\pi G_4(\rho + p) \ ,\quad
\rho + p= {4 Ml \over 3 V_3 a^4}\ .
\eea
Here 4d Newton constant $G_4$ is given by
\be
\label{e12}
G_4={2G \over l}\ .
\ee

For SAdS, The entropy and the thermodynamical energy have 
the following form:
\bea
\label{ent2}
{\cal S }&=& {V_{3}\pi r_H^3 \over 4\pi G}\ ,\\
\label{ener2}
E&=&M\ .
\eea
Here $r_H$ is the radius of the BH horizon. 
By comparing (\ref{ener2}) and (\ref{e10}), we find 
\be
\label{AAA1}
E={l \over a}E_4\ ,\quad E_4= \rho V_3 a^3\ .
\ee
Note that when $a$ is large, the metric 
(\ref{AdSS}) has the following form:
\be
\label{eq13} 
ds_{\rm AdS-S}^2 \rightarrow {a^2 \over l^2}\left(-dt^2 
+ l^2 \sum_{i,j}^{d-1} 
g_{ij}dx^i dx^j\right)\ ,
\ee
which tells that the time $\tilde t$ on the brane is equal 
to the AdS time $t$ times the factor ${a \over l}$:
\be
\label{eq14}
t_{\rm CFT}={a \over l}t\ .
\ee
Therefore Eq.(\ref{AAA1}) expresses that 
the energy on the brane is related 
with the energy $E$ in AdS by a factor ${l \over a}$ \cite{SV}. 

By using (\ref{e10}) and (\ref{e11}), we find 
\be
\label{trace1}
0=-\rho + 3p\ ,
\ee
which tells that the trace of the energy-stress tensor coming 
from the matter on the brane vanishes:
\be
\label{trace2}
{T^{{\rm matter}\ \mu}}_\mu=0\ .
\ee
Therefore the matter on the brane can be 
regarded as the radiation, i.e., the massless fields. 
In other words, field theory on the brane should be conformal one. 

In \cite{EV}, it was shown that the FRW equation in 
$d$-dimensions can be regarded as a $d$-dimensional analogue of
the Cardy formula of 2d conformal field theory (CFT):
\be
\label{CV1}
\tilde {\cal S}=2\pi \sqrt{
{c \over 6}\left(L_0 - {k \over d-2}{c \over 24}\right)}\ .
\ee
In the present case for $d=4$ case, identifying
\bea
\label{CV2}
{2\pi E_4 a \over 3} &\Rightarrow& 2\pi L_0 \ ,\nn
{(V \over 8\pi G a} &\Rightarrow& {c \over 24} \ ,\nn
{HV \over 2G} &\Rightarrow& \tilde {\cal S}\ ,
\eea
the FRW-like equation (\ref{e10}) has the form (\ref{CV1}). 

The total entropy of the universe could be conserved in the 
expansion. Then one can evaluate holographic (Hubble) entropy 
$\tilde{\cal S}$ in (\ref{CV2}) 
when the brane crosses the horizon $r=r_H$. When $r=r_H$, 
Eq.(\ref{e10}) tells that 
\be
\label{CV3}
H=\pm {1 \over l}\ .
\ee
Here the plus sign corresponds to the expanding brane universe 
and the minus one to the contracting universe.  Taking 
the expanding case and using (\ref{CV2}), we find
\be
\label{CV4}
\tilde{\cal S}={r_H^3 V_3 \over 4G}\ ,
\ee
which is nothing but the black hole entropy in (\ref{ent2}). 
Then the expression (\ref{HS}) can be naturally understood from the 
context of the brane world.

\end{document}